\documentclass{iau} 
\usepackage{graphicx}

\title[GWimagefiltering] 
{Improving agnostic searches of Gravitational Waves from Neutron Star instabilities using image filtering}

\author[Lorenzo Pierini]   
{Lorenzo Pierini}

\affiliation{Dipartimento di Fisica, Università degli Studi di Roma La Sapienza, \\
	[\affilskip]
Istituto Nazionale di Fisica Nucleare, Sezione Roma 1, \\
Piazzale Aldo Moro 2, 00185, Rome, Italy \\email: {\tt lorenzo.pierini@roma1.infn.it}}

\pubyear{2021}
\volume{363}  
\jname{Neutron Star Astrophysics at the Crossroads:\\
Magnetars and the Multimessenger Revolution}
\editors{E. Troja \& M. Baring, eds.}
\begin{document}

\maketitle

\begin{abstract}
In this paper I present a method to enhance the search sensitivity for long transient Gravitational Waves produced by Neutron Star instabilities. This method consists in a selective image filter, called Triangular Filter, to be applied to data spectrograms. It is shown that thanks to this implementation a $\sim20\%$ gain in sensitivity is achievable.
\end{abstract}

\firstsection 
\section{Introduction}

The search of Gravitational Waves (GW) emitted by isolated or binary Neutron Stars (NS) is one of the biggest challenges in GW astronomy. In this work I focus on transient GWs - with typical timescales of $\mathcal{O}$($10^3$)s - produced by NS instabilities, see \cite{lasky} for a review. Such instabilities can develop in young supernova remnants or in the post-merger remnant of binary NS inspirals, for which an electromagnetic counterpart is expected. We can then assume to know the source position fairly well, while rotational parameters are largely unknown. The resulting huge parameter space makes matched filter searches computationally unfeasible and different strategies have to be used. One example is the \textit{Generalized Frequency Hough} (GFH), described in \cite{GFH} and used the first time in \cite{search}. This procedure works on spectrograms of the detectors data: it searches for time-frequency patterns of the expected GW signal, assumed to follow a power law equation. This is done transforming data into the signal parameter space though the GFH transform. This procedure is computationally feasible and is also more robust with respect to deviations from theoretical models, but it pays a price in terms of sensitivity loss with respect to matched filter techniques. The goal of my work is to reduce this sensitivity gap.
\section{Method: Triangular Filter on spectrograms}
\begin{figure*}
	\includegraphics[height=4.8cm]{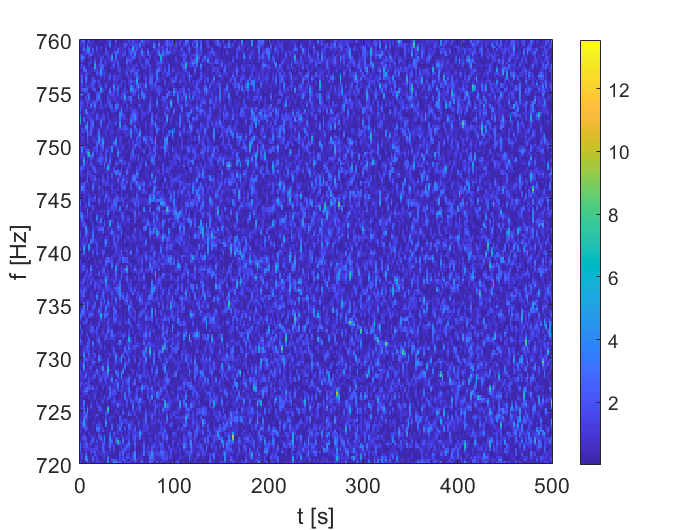}
	\includegraphics[height=4.8cm]{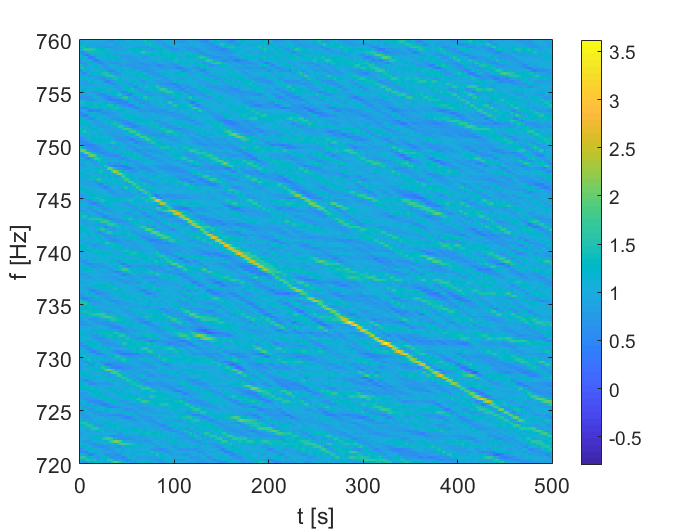}
	\caption{Effect of the Triangular Filter application on spectrograms. On the left: whitened spectrogram of LIGO Livingston O2 data containing a simulated long transient GW starting at 750 Hz. The signal is visible but its energy does not exceed the mean noise level. On the right: the same map after filtering. Here the signal has become an outlier with respect to the noise.}
	\label{fig:effect}
\end{figure*}

The method I developed works directly on the spectrograms, where we search for GW time-frequency patterns, and consists of an image filter to enhance the signal-to-noise ratio (SNR). This filter is applied in the 2D Fourier domain, and it is suited to filter out signals with decreasing frequency patterns. Its computational cost is negligible, since it is applied just one time before appying the multiple GFH transforms. It is called \textit{Triangular Filter} because of its shape in the 2D Fourier space. A decreasing time-frequency pattern is characterized by negative slope tangents: their information is stored in a well defined region of the 2D Fourier space, so the filter selects only the regions of the Fourier Transform of the spectrogram that correspond to signals slowing down within a given range. A visual proof  of the filtering effect is shown in Figure \ref{fig:effect}, where the color scale reveals the enhanced SNR after the filter application. Repeating the simulation with randomly localized sources and varying signal amplitudes, it is possible to compute the efficiency curves in the standard case of GFH and after applying the Triangular Filter. If we set a fiduciary efficiency level at $95\%$, it turns out that we can obtain a sensitivity gain of $\sim 20\%$.
\begin{figure*}
	\centering
	\includegraphics[height=4.4cm]{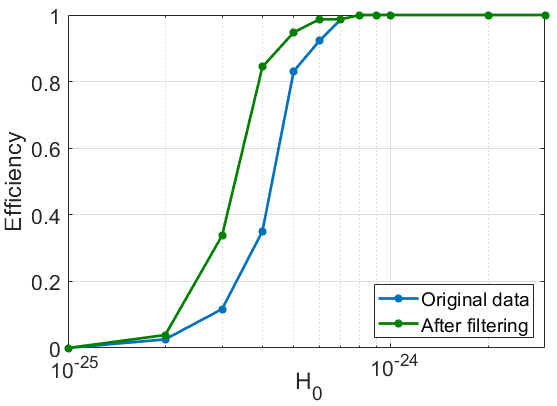}
	\caption{Efficiency curves obtained through multiple injections of simulated GW r-mode signals - see \cite{lasky} - on LIGO Livingston O2 data starting from GPS=1253179440s and with initial frequency in the range [700-770]Hz. The blue line represents the curve obtained with the ordinary GFH procedure, whereas the green one is obtained using the Triangular Filter.}
	\label{fig:eff}
\end{figure*}
\section{Implications and future developments}
The results obtained show clearly the sensitivity gain that can be achieved thanks to the Triangular Filter technique. The same filter can be implemented to search for signals with growing or wandering frequency, just restricting to frequency derivatives not greater than the optimal resolution of the spectrogram. In the future, this filter will be implemented as a crucial step of a new search procedure, in order to enhance the detection probability. Another interesting search topic will be to evaluate the impact of this filter when it is applied as a pre-processing step on deep learning based methods.

\end{document}